\documentclass[journal]{IEEEtran}
\usepackage{amsmath,amsfonts}
\usepackage{algorithmic}
\usepackage{array}
\usepackage[caption=false,font=normalsize,labelfont=sf,textfont=sf]{subfig}
\usepackage{textcomp}
\usepackage{stfloats}
\usepackage{url}
\usepackage{verbatim}
\usepackage{graphicx}
\def\BibTeX{{\rm B\kern-.05em{\sc i\kern-.025em b}\kern-.08em
    T\kern-.1667em\lower.7ex\hbox{E}\kern-.125emX}}
\usepackage{balance}
\usepackage{life-hts}
\usepackage{bm}
\usepackage{booktabs}
\usepackage{upgreek}
\usepackage{multirow}
\begin{document}
\title{Simultaneous Multi-Scale Homogeneous \\ H-Phi Thin-Shell Model for Efficient Simulations \\ of Stacked HTS Coils}
\author{Louis Denis, Benoît Vanderheyden, and Christophe Geuzaine
\thanks{L. Denis is a research fellow of the Fonds de la Recherche Scientifique - FNRS. (\textit{Corresponding author: L. Denis})

The authors are with the Department of Electrical Engineering and Computer Science, Institut Montefiore B28 in the University of Liege, 4000 Liege, Belgium (e-mail: louis.denis@uliege.be).}}

\markboth{October 2025}
{L. Denis, B. Vanderheyden, and C. Geuzaine: Simultaneous Multi-Scale Homogeneous H-Phi Thin-Shell Model for Efficient Simulations of Stacked HTS Coils}

\makeatletter
\def\ps@IEEEtitlepagestyle{
  \def\@oddfoot{\mycopyrightnotice}
  \def\@evenfoot{}
}
\def\mycopyrightnotice{
  {\footnotesize
  \begin{minipage}{\textwidth}
    \fbox{\parbox{\textwidth}{
        This work has been submitted to a journal for possible publication. Copyright may be transferred without notice, after which this version may no longer be accessible.
        }}
  \end{minipage}
  }
}

\maketitle

\begin{abstract}
The simulation of large-scale high-temperature superconducting (HTS) magnets is a computational challenge due to the multiple spatial scales involved, from the magnet to the detailed turn-to-turn geometry. To reduce the computational cost associated with finite-element (FE) simulations of insulated HTS coils, the simultaneous multi-scale homogeneous (SMSH) method can be considered. It combines a macroscopic-scale homogenized magnet model with multiple single-tape models and solves both scales monolithically. In this work, the SMSH method is reformulated using the $h$-$\phi$ thin-shell (TS) approximation, where analyzed tapes are collapsed into thin surfaces, simplifying mesh generation. Moreover, the magnetic field is expressed as the gradient of the magnetic scalar potential outside the analyzed tapes. The discretized field is then described with nodal functions, further reducing the size of the FE problem compared to standard $h$ formulations. The proposed $h$-$\phi$ SMSH-TS method is verified against state-of-the-art homogenization methods on a \mbox{2-D} benchmark problem of stacks of HTS tapes. The results show good agreement in terms of AC losses, turn voltage and local current density, with a significant reduction in simulation time compared to reference models. All models are open-source.
\end{abstract}


\section{Introduction}
\IEEEPARstart{R}{eliable} predictions of AC losses in high-temperature superconducting (HTS) magnets are essential for the development of next-generation large-scale applications~\cite{Coombs2024}, such as fusion magnets. However, finite-element (FE) simulations of HTS magnets remain particularly challenging due to the multi-scale nature of the problem, the nonlinear and anisotropic magnetic response of HTS coated conductors (or \textit{tapes}), and their large aspect ratios~\cite{Sirois2015}.

Homogenization techniques~\cite{ElFeddi1997a} can help in addressing these challenges, as they reduce the size of the numerical problem by solving it for a medium with averaged magnetic properties. Accordingly, stacks of HTS tapes are represented as homogenized bulks with equivalent anisotropic properties. This leads to a significant computational speedup compared to detailed models, as studied in several works~\cite{Zermeno2013a,Zermeno2014c,Berrospe-Juarez2019,Vargas-Llanos2022,Santos2024a}, using the $h$, $t$-$a$, and $j$-$a$ FE formulations. Recent extensions include the coupling with thermal physics~\cite{Klop2024,Dadhich2024b}, as well as the \textit{foil winding} (FW) homogenization technique~\cite{Paakkunainen2025,Denis2025} that recovers the voltage distribution across individual turns within the stacks of HTS tapes.

Another technique to reduce the size of the problem is the multi-scale approach that relies on the coupling of a macroscopic-scale magnet model with multiple mesoscopic-scale models of single conductors~\cite{Queval2016a,Berrospe-Juarez2018,Wang2023a}. Thus, AC losses are computed on the basis of the local fields at the scale of each HTS tape. However, this requires an iterative communication between scales, which leads to reasonable simulation times only on massively parallel computers, where the mesoscopic models can be solved concurrently~\mbox{\cite{Berrospe-Juarez2021,Denis2025a,Niyonzima2016}}.

\begin{figure}[!t]
\centering
\includegraphics[width=0.95\columnwidth]{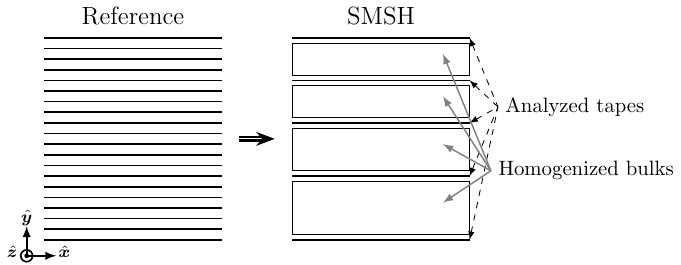}
\caption{Detailed stack of $N_{\text{c}}$ tapes (left), with its equivalent in the SMSH model, which replaces non-analyzed tapes with homogenized bulks (right).}
\label{fig:principle-intro}
\end{figure}

To overcome these limitations, the simultaneous multi-scale homogeneous (SMSH) method~\cite{Berrospe-Juarez2021} was introduced as an alternative approach. As represented in Fig.~\ref{fig:principle-intro}, it combines the detailed resolution of a selected set of \textit{analyzed tapes} with the homogenization of the remaining tapes into bulks. This allows the macroscopic and mesoscopic scales to be solved monolithically, i.e. with a single numerical resolution. The SMSH method has been developed with the total $h$ and $t$-$a$ FE formulations in~\cite{Berrospe-Juarez2021}, and a refined approach based on the $t$-$a$ formulation was later proposed in~\cite{Wang2022,Wang2025a}.

Here, the SMSH method is extended to the $h$-$\phi$ FE formulation, in which the magnetic field outside analyzed tapes is described via the magnetic scalar potential $\phi$. Analyzed tapes are modelled with the $h$-$\phi$ thin-shell (TS) approximation~\cite{deSousaAlves2021,deSousaAlves2022}. The tapes are collapsed into surfaces to simplify mesh generation and further increase the computational efficiency. Unlike its $t$-$a$~\cite{Zhang2016} and $h$-$a$~\cite{Bortot2020} counterparts, the $h$-$\phi$ TS model involves the magnetic scalar potential outside the tapes, with a much smaller number of degrees of freedom in 3-D geometries~\cite{deSousaAlves2022}. Moreover, it has been successfully extended to 3-D magneto-thermal simulations of HTS coils in~\cite{Schnaubelt2023b,Schnaubelt2024c}.

This work is organised as follows: Section~\ref{sec:2} describes the proposed $h$-$\phi$ SMSH-TS model, combining the SMSH method with the $h$-$\phi$ TS model. Section~\ref{sec:3} introduces the 2-D benchmark problem, while the SMSH-TS approach is verified and compared to both detailed and state-of-the-art homogenized models in Section~\ref{sec:4}.

\section{Finite-element formulations \label{sec:2}}
Before describing the core of the proposed SMSH-TS model, the conventional $h$-$\phi$ FE formulation is recalled and applied to stacks of HTS
tapes.

\subsection{Conventional $h$-$\phi$ FE formulation \label{sec:ref-hphi}}
The reference model is constructed for a stack of $N_{\text{c}}$ insulated HTS tapes, forming the conducting subset $\Oc = \bigcup_{i=1}^{N_{\text{c}}} \Omega_{\text{c},i}$ of the computational domain $\O$. Turns are arranged in series and the net current in each tape is $I_{\text{t}}$. The superconducting layer of each HTS tape is represented with its real thickness. The $h$-$\phi$ FE formulation, as detailed in~\cite{Dular2020a}, aims to find the magnetic field $\h \in \mathcal{H}_I(\text{curl},\Omega)$ such that
\begin{equation}
    \volInt{\mu\partial_t\h}{\h'}{\O} + \volInt{\rho\,\curl{\h}}{\curl{\h}'}{\Oc} = 0 \label{eq:ref-hphi}
\end{equation}
holds $\forall \h' \in \mathcal{H}_0(\text{curl},\Omega)$, with $\mu=\mu_0$ the magnetic permeability and $\rho$~the electric resistivity. Here, the shorthand $\volInt{\cdot}{\cdot}{\O}$ denotes the volume integral over $\O$ of the inner product of its arguments. The vector space of square-integrable vector fields with square-integrable curl that satisfy strong (resp. vanishing) current constraints in $\Oc$ is represented by $\mathcal{H}_I(\text{curl},\Omega)$ (resp. $\mathcal{H}_0(\text{curl},\Omega)$). The HTS resistivity is described by the power-law (PL) \cite{Kim1962,Anderson1962}:
\begin{equation}
    \rho(\j,\b) = \frac{e_{\text{c}}}{j_{\text{c}}(\b)} \left(\frac{\lVert \j \rVert}{j_{\text{c}}(\b)} \right)^{n-1}, \label{eq:power-law}
\end{equation}
with $\j = \curl \h$ the current density, $\b = \mu_0 \h$ the magnetic flux density, $\jc$ the HTS critical current density, and $e_{\text{c}} = 10^{-4}$~V/m the critical electric field.

In the non-conducting domain $\Occ = \O \setminus \Oc$, the magnetic field is curl-free and can be expressed as $\h = -\grad \phi$. Thus, the magnetic field is discretized as 
\begin{equation}
    \h = \sum_{e \in \mathcal{E}(\Oc \setminus \partial \Oc)} h_e\,\vec\psi_e + \sum_{n \in \mathcal{N}(\Occ)} \phi_n\,\grad\psi_n + \sum_{i=1}^{\Nc} I_{\text{t}} \,\bm c_{i}, \label{eq:hphi-discrete}
\end{equation}
with $\vec \psi_e$ (resp. $\psi_n$) edge (resp. nodal) shape functions and $\bm c_{\text{i}}$ the edge cohomology basis functions \cite{Pellikka2013} (or \textit{cuts}) associated to the conducting subdomain $\O_{\text{c},i}$. From~\eqref{eq:hphi-discrete} onwards, discretized physical fields are referred to by their continuous notation, for conciseness. The discretization~\eqref{eq:hphi-discrete} results in fewer degrees of freedom (DoFs) than the total $h$ FE formulation relying exclusively on edge shape functions. Notably, the turn voltage $V_i$ in conductor $\Omega_{\text{c},i}$ can be evaluated with a single equation~\cite{Dular2020a}:
\begin{equation}
    V_i = - \volInt{\mu\partial_t\h}{\bm c_{i}'}{\O} - \volInt{\rho\,\curl{\h}}{\curl{\bm c_{i}}'}{\Oc}, \label{eq:v-computation}
\end{equation}
with $\bm c_{i}'$ the test function associated to the corresponding cohomology basis function $\bm c_{i}$. This relation also holds in the TS model described below.

\subsection{Thin-shell $h$-$\phi$ FE formulation \label{sec:ts-hphi}}

The $h$-$\phi$ TS model, introduced in~\cite{deSousaAlves2021}, reduces each thin conducting layer $\O_{\text{c},i}$ to a surface $\Gamma_{\text{c},i}$. Accordingly, the second term in the conventional $h$-$\phi$ FE formulation~\eqref{eq:ref-hphi} must be replaced by surface integrals on $\Gamma_{\text{c}} = \bigcup_{i=1}^{\Nc} \Gamma_{\text{c},i}$, see~\cite{deSousaAlves2021}. 

\begin{figure}[!t]
\centering
\includegraphics[width=0.95\columnwidth]{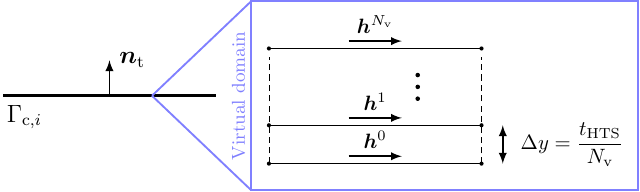}
\caption{Internal representation of the $N_{\text{v}}$ virtual elements inside one thin-shell $\Gamma_{\text{c},i}$, each of virtual height $\Delta y$. Adapted from~\cite{deSousaAlves2021}.}
\label{fig:sketch-ts}
\end{figure}

The TS model introduces a discontinuity in $\h$ across the collapsed tape. The corresponding discontinuity in $\phi$ is here enforced by introducing a crack at the mesh level, similarly to~\cite{deSousaAlves2021}. Within the thin-shell itself, $N_{\text{v}}$ virtual elements are introduced to describe the magnetic field penetration across the tape thickness $t_{\text{HTS}}$. As shown in Fig~\ref{fig:sketch-ts}, this involves $(N_{\text{v}}+1)$ auxiliary fields $\h^{k}$ in the TS, with the magnetic field tangential to the top and bottom tape surfaces denoted $\h^{N_{\text{v}}}$ and $\h^0$, respectively. The current density, averaged over the TS thickness, is
\begin{equation}
    \j = \bm{n}_{\text{t}} \times \left( \frac{\h^{N_{\text{v}}} - \h^0}{t_{\text{HTS}}} \right), \label{eq:j-TS}
\end{equation}
with $\bm{n}_{\text{t}}$ the normal to the top surface of the TS. Instantaneous AC losses in $\Gamma_{\text{c}}$ can be computed~\cite{deSousaAlves2021} as
\begin{equation}
    q = \sum_{k=1}^{N_{\text{v}}}\frac{N_{\text{v}}}{t_{\text{HTS}}}\int_{\Gamma_{\text{c}}} \rho~\left\lVert \h^{k} - \h^{k-1} \right\rVert^2~d\Gamma, \label{eq:losses-TS}
\end{equation}
where $\rho$ is modelled by the power-law function~\eqref{eq:power-law}.

In $\O \setminus \Gamma_{\text{c}}$, the magnetic field satisfies $\h = - \grad \phi$ and is thus discretized without the first term in~\eqref{eq:hphi-discrete}. Inside the thin-shells, auxiliary fields $\h^k$ are discretized with edge shape functions tangential to the surface, while $\h$ is interpolated with first-order Lagrange polynomials across the virtual thickness~\cite{deSousaAlves2021}. Note that the normal component of $\h$ is not defined inside the TS. It is here recovered as the average trace of $\h \cdot \bm{n}_{\text{t}}$ on the top and bottom surfaces of the thin-shell.

\subsection{SMSH-TS $h$-$\phi$ FE formulation \label{sec:smsh-ts}}

\begin{figure}[!t]
\centering
\includegraphics[width=0.95\columnwidth]{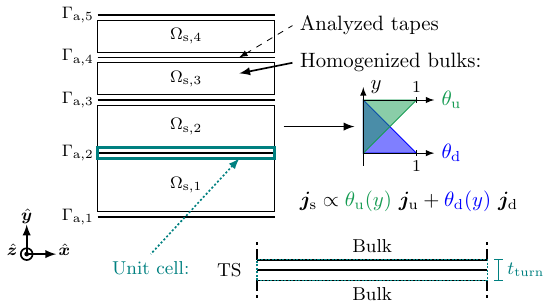}
\caption{Principle and notations of the SMSH-TS model, replacing non-analyzed tapes with homogenized bulks, in which the current density $\bm{j}_{\text{s}}$ is interpolated from neighbouring analyzed tapes (top). Focus on the unit cell around single analyzed tapes, represented with the $h$-$\phi$ TS model (bottom).}
\label{fig:principle}
\end{figure}

The proposed $h$-$\phi$ SMSH-TS model is illustrated in Fig.~\ref{fig:principle}. It shares features with the \textit{simultaneous multi-scale homogeneous} model in~\cite{Berrospe-Juarez2021} and is adapted here to the $h$-$\phi$ TS formulation. A subset of \textit{analyzed tapes} $\Gamma_{\text{a},i}$ is considered, with $\Gamma_{\text{a}} = \bigcup_{i=1}^{N_{\text{a}}} \Gamma_{\text{a},i} \subset \Gamma_{\text{c}}$, and with a current density obtained from the previously described $h$-$\phi$ TS formulation. Note that the efficiency of the multi-scale approach relies on a thoughtful selection of the analyzed tapes before running the simulation, see~\cite{Queval2016a,Berrospe-Juarez2018}. The TS formulation is used to provide detailed AC loss estimations~\eqref{eq:losses-TS} in the analyzed tapes. For non-analyzed tapes, losses are obtained through the piecewise cubic Hermite interpolating polynomial (PCHIP) method~\cite{Berrospe-Juarez2018,Berrospe-Juarez2021}.

Between two successive analyzed tapes $\Gamma_{\text{a},i}$ and $\Gamma_{\text{a},i+1}$, the $N_{\text{s},i}$ non-analyzed tapes are merged into a homogenized bulk $\Omega_{\text{s},i}$. Its effective thickness is $N_{\text{s},i} \cdot t_{\text{turn}}$, with $t_{\text{turn}}$ the thickness of a complete turn. By symmetry, this configuration defines a unit cell with airgaps around each analyzed tape, as illustrated in Fig.~\ref{fig:principle}. Such a unit cell locally approximates the actual geometry of the analyzed tapes and their direct environment and reproduces the local stray fields seen by analyzed tapes with a fair accuracy. Note that unlike~\cite{Berrospe-Juarez2021}, the non-analyzed tapes adjacent to analyzed tapes are included in the homogenized bulks. 

In each bulk $\Omega_{\text{s},i}$, the current density is not computed explicitly from a FE resolution, but is interpolated from the neighbouring analyzed tapes. Introducing the superconducting filling factor $f_{\text{SC}}=t_{\text{HTS}}/t_{\text{turn}}$, the resulting \textit{source} current density is obtained with
\begin{equation}
    \bm{j}_{\text{s}} = f_{\text{SC}} \left( \theta_{\text{u}}(y)~\bm{j}_{\text{u}} + \theta_{\text{d}}(y)~\bm{j}_{\text{d}} \right), \label{eq:js-def}
\end{equation}
in which $\bm{j}_{\text{u}}$ and $\bm{j}_{\text{d}}$ denote the current density computed with~\eqref{eq:j-TS} in analyzed tapes $\Gamma_{\text{a},i+1}$ and $\Gamma_{\text{a},i}$, respectively. The interpolation is performed with the first-order Lagrange polynomials $\theta_{\text{d}}$ and $\theta_{\text{u}}$ depicted in Fig.~\ref{fig:principle}.

Consequently, the magnetic field is no longer curl-free in the \textit{source} domain $\Omega_{\text{s}} = \bigcup_{i=1}^{N_{\text{a}}-1} \Omega_{\text{s},i}$. This is taken into account in the $h$-$\phi$ FE formulation by decomposing the magnetic field~\cite{Dular1994} into
\begin{equation}
    \h = \h_{\text{s}} + \h_{\text{r}}, \label{eq:h-decomposition}
\end{equation}
with $\curl \h_{\text{s}} = \j_{\text{s}}$ in $\O_{\text{s}}$ and $\curl \h_{\text{r}} = \bm{0}$ in $\O \setminus \Gamma_{\text{a}}$. This allows the \textit{reaction} magnetic field $\h_{\text{r}}$ to be computed with the $h$-$\phi$ TS formulation, since $\h_{\text{r}} = - \grad \phi$ in $\O \setminus \Gamma_{\text{a}}$, provided~\eqref{eq:h-decomposition} is set in the first term of~\eqref{eq:ref-hphi}. Once again, the field $\h_{\text{r}}$ is discretized without the first term in~\eqref{eq:hphi-discrete}.

The \textit{source} magnetic field $\h_{\text{s}}$ is computed with a weak projection~\cite{Geuzaine1999} that aims to find $\h_{\text{s}} \in \mathcal{H}_{I_{\text{s}}}(\text{curl},\O_{\text{s}})$ such that
\begin{equation}
    \volInt{\curl\h_{\text{s}}}{\curl\h_{\text{s}}'}{\O_{\text{s}}} = \volInt{\j_{\text{s}}}{\curl\h_{\text{s}}'}{\O_{\text{s}}},
    \label{eq:hs-weak-projection}
\end{equation}
holds $\forall \h_{\text{s}}' \in \mathcal{H}_0(\text{curl},\Omega)$, with $\j_{\text{s}}$ given by~\eqref{eq:js-def}. The field $\h_{\text{s}}$ is discretized as
\begin{equation}
    \h_{\text{s}} = \sum_{e \in \mathcal{E}(\O_{\text{s}}\setminus \mathcal{T}_{\text{s}})} h_{\text{s},e}\,\vec\psi_e + \sum_{i=1}^{N_{\text{a}}-1} I_{\text{s},i} \,\bm c_{\text{s},i}, \label{eq:hs-discrete}
\end{equation}
with $I_{\text{s},i} = N_{\text{s},i} \cdot I_{\text{t}}$ the net current carried by the non-analyzed tapes merged into $\O_{\text{s},i}$, consistent with the prescribed $I_{\text{t}}$. The gauge of $\h_{\text{s}}$ is fixed by strongly enforcing its restriction to the edges of a co-tree built in $\O_{\text{s}}$, with the tree $\mathcal{T}_{\text{s}}$ being complete on $\partial \O_{\text{s}}$~\cite{Geuzaine1999}. Notably, it is numerically convenient to generate the source cohomology basis functions $\bm c_{\text{s},i}$ without intersecting $\Gamma_{\text{a}}$, thus enforcing $\h = \h_{\text{r}}$ in $\Gamma_{\text{a}}$. This ensures that the surface terms arising from the $h$-$\phi$ TS formulation~\cite{deSousaAlves2021} in~\eqref{eq:ref-hphi} can be used without any change.

\begin{figure}[!t]
\centering
\includegraphics[width=0.6\columnwidth]{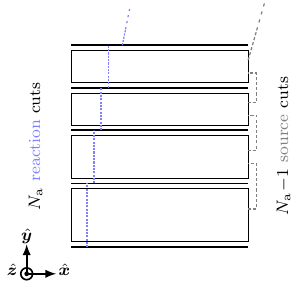}
\caption{Conceptual sketch of the cohomology basis functions (or \textit{cuts}) involved in the discretizations of the reaction~\eqref{eq:hphi-discrete} and source magnetic fields~\eqref{eq:hs-discrete}.}
\label{fig:cohomology}
\end{figure}

To summarize, the proposed $h$-$\phi$ SMSH-TS FE model is obtained by combining the weak formulations~\eqref{eq:ref-hphi} and~\eqref{eq:hs-weak-projection}, including additional terms from the $h$-$\phi$ TS formulation detailed in~\cite{deSousaAlves2021}, with strongly satisfied relationships~\eqref{eq:j-TS},~\eqref{eq:js-def} and~\eqref{eq:h-decomposition}. The cohomology basis functions in discretizations~\eqref{eq:hphi-discrete} and~\eqref{eq:hs-discrete}, illustrated in Fig.~\ref{fig:cohomology}, are created at the mesh level with the Gmsh open-source software~\cite{Geuzaine2009}. 

While the method has been described for a single stack of HTS tapes, its extension to multiple stacks is straightforward. Note that the contribution of normal conducting layers across each tape can also be included in the $h$-$\phi$ TS model~\cite{deSousaAlves2021}.

\section{Benchmark problem description \label{sec:3}}
\begin{figure}[!t]
\centering
\includegraphics[width=0.9\columnwidth]{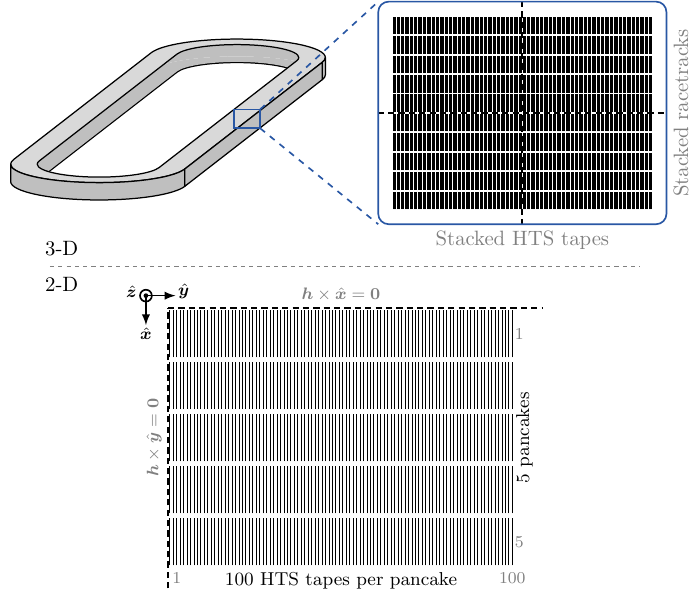}
\caption{Three-dimensional model of the racetrack coil to be studied (top), together with the simplified quarter cross-section modeled for verification (bottom). Adapted from~\cite{Berrospe-Juarez2021}.}
\label{fig:benchmark-geometry}
\end{figure}

The considered benchmark problem has been extensively investigated in~\cite{Berrospe-Juarez2019,Queval2016a,Berrospe-Juarez2021,Wang2022}. It corresponds to the 2-D cross-section of the straight section of a racetrack coil, made of ten pancakes with 200 turns each. By symmetry, one quarter of the cross-section is simulated as depicted in Fig.~\ref{fig:benchmark-geometry}. The boundary conditions impose vanishing tangential magnetic fields on the symmetry planes.

The field-dependent critical current density $\jc(\b)$ of the HTS layer is described by a Kim-like model~\cite{Kim1962}:
\begin{equation}
    \jc(\b) = \frac{j_{\text{c}0}}{\left(1 + \frac{\sqrt{k_{\text{c}}^2 b_x^2 + b_y^2}}{b_0}\right)^{\alpha}}, \label{eq:jcb}
\end{equation}
where $j_{\text{c}0}$, $k_{\text{c}}$, $b_0$, and $\alpha$ are material parameters. The geometrical and material parameters used in simulations are gathered in Table~\ref{tab:numerical-parameters}.

\begin{table}[t!]
    \vspace*{-.5em}
    \begin{center}
    \caption{Numerical parameters of the study case.}
    \label{tab:numerical-parameters}
    \vspace*{-0.5em}
    \begin{tabular}{cc|cc}
    \toprule
    Parameter & Value & Parameter & Value\\
    \midrule
    $j_{\textrm{c}0}$ & $28$~kA/mm$^2$ & HTS layer width & $4~\text{mm}$\\
    $b_0$ & $42.65$~mT & HTS thickness & $1~\upmu\text{m}$ \\
    $k_{\text{c}}$ & $0.29515$ & Turn width & $4.4~\text{mm}$ \\
    $\alpha$ & $0.7$ & Turn thickness & $293~\upmu\text{m}$\\
    PL $n$-index & $38$ & Transport current & $11$~A, $50$~Hz\\
    \bottomrule
    \end{tabular}
    \end{center}
    \vspace*{-1em}
\end{table}

The proposed $h$-$\phi$ SMSH-TS model is assessed against both detailed and homogenized models.

The detailed models explicitly represent the 500 HTS layers. The following formulations are considered: $h$-$\phi$ (cf. section~\ref{sec:ref-hphi}), total $h$ (referred to as the $h$-formulation in~\mbox{\cite{Queval2016a,Berrospe-Juarez2021}}), $h$-$\phi$ TS (cf. section~\ref{sec:ts-hphi}) and $t$-$a$~\cite{Zhang2016}. The $h$-$\phi$ formulation is hereafter referred to as the \textit{reference} model. 

Two $h$-based homogenization techniques are considered: the homogenized model from~\cite{Zermeno2013a}, here adapted to the $h$-$\phi$ formulation (\textit{$h$-$\phi$ Vanilla}), as well as the $h$-$\phi$ foil-winding ($h$-$\phi$ FW) model~\cite{Denis2025}. Also, the $t$-$a$ SMSH model is implemented, for direct comparison with the proposed approach. It corresponds to the $t$-$a$ \textit{simultaneous multi-scale homogeneous} in~\cite{Berrospe-Juarez2021}, with the inclusion of non-analyzed tapes adjacent to analyzed tapes within bulks.

All models are discretized with 50 equidistant mesh elements along the HTS width. In the conductors, detailed $h$-$\phi$ and total $h$ models, as well as the homogenized models, use rectangular elements. All detailed models use a single element through the HTS thickness, with $N_{\text{v}}=1$ in the $h$-$\phi$ TS model. Both SMSH models consider seven analyzed tapes per pancake (indices ${1,25,66,88,96,99,100}$). Each homogenized bulk is discretized with up to three elements through the thickness depending on its size, while the other $h$-$\phi$-based homogenized models use 12 equidistant elements through the stack thickness. A single period of applied transport current is simulated with 600 constant time steps.

Two quantities of interest are defined for model comparison. The first quantity provides a global metric for comparing the models: the AC loss $P$ averaged over the second half-cycle, defined as
\begin{equation}
    P = \frac{2}{T} \int_{T/2}^{T} q(t)~dt, \label{eq:P-average}
\end{equation}
with $q$ the instantaneous AC loss and $T$ the current cycle period. The corresponding relative error with respect to the reference is denoted $e_{P}$. The second quantity is sensitive to local variations among the models: the $R^2$ coefficient of determination of the current density distribution, defined as
\begin{equation}
    R^2 = 1 - \frac{\int_{0}^{T}\int_{\Omega_{\text{HTS}}} (\j - \j_{\text{ref}})^2~d\Omega_{\text{HTS}}~dt}{\int_{0}^{T}\int_{\Omega_{\text{HTS}}} (\bar{\j}_{\text{ref}} - \j_{\text{ref}})^2~d\Omega_{\text{HTS}}~dt}, \label{eq:j-R2}
\end{equation}
with $\Omega_{\text{HTS}}$ the generic domain of all 500 HTS tapes, $\j_{\text{ref}}$ the current density from the reference $h$-$\phi$ model and $\bar{\j}_{\text{ref}}$ its mean value.

All models are implemented in the open-source FE solver GetDP~\cite{Dular1998}, while meshes are generated with Gmsh~\cite{Geuzaine2009}. All source files are available online.\footnote{[Online]. Available: \url{www.life-hts.uliege.be}.}

\section{Numerical results \label{sec:4}}
\begin{figure}[!t]
\centering
\includegraphics{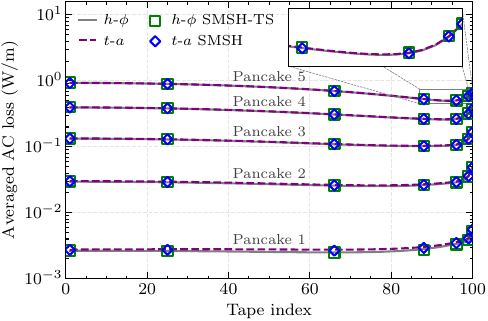}
\caption{Averaged AC loss per turn, computed with detailed and SMSH models.}
\label{fig:loss_v_position}
\end{figure}

Figure \ref{fig:loss_v_position} demonstrates that the proposed $h$–$\phi$ SMSH-TS model accurately reproduces the averaged AC losses in analyzed tapes, with results in excellent agreement with both detailed $h$–$\phi$ and $t$–$a$ models. The overlap with the $t$–$a$ SMSH results highlights the equivalence of the two formulations.

\begin{figure}[!t]
\centering
\includegraphics{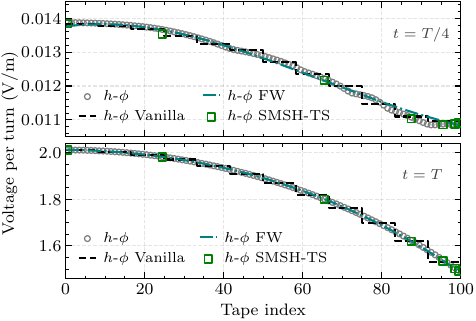}
\caption{Voltage per turn in pancake 5, evaluated at the first current peak (top) and after one current cycle (bottom), computed with various $h$-$\phi$ models.}
\label{fig:voltage_v_position}
\end{figure}

The developed model also provides the turn voltage distribution across tapes, as shown in Fig.~\ref{fig:voltage_v_position}. Despite negligible discrepancies at the first current peak (when the inductive voltage is zero), the agreement with both detailed and homogenized $h$–$\phi$ models remains excellent. Notably, the $h$-$\phi$ FW model, which approximates the voltage distribution with a third-order global polynomial~\cite{Denis2025}, smooths out the staircase shape observed with the homogenized $h$-$\phi$ Vanilla model.

\begin{table}[t!]
    \vspace*{-.5em}
    \begin{center}
    \caption{Comparison of different models.}
    \label{tab:numerical-performance}
    \vspace*{-0.5em}
        \begin{tabular}{c|cccccc}
            \toprule
            \multirow{2}{*}{Model} & DoFs & $P$ & $e_P$ & $R^2$ & $t_{\text{a}}^*$ & $t_{\text{s}}^*$ \\
             & - & W/m & - & - & min & min \\
            \midrule
            $h$-$\phi$ & 192k & 124.8 & - & - & 90 & 99 \\
            total $h$ & 357k & 124.8 &0.003\% & 1-10$^{\text{-7}}$ & 28 & 180 \\
            $h$-$\phi$ TS & 172k & 125.7 & 0.75\% & 0.988 & 91 & 89 \\
            $t$-$a$ & 148k & 125.4 & 0.45\% & 0.990 & 39 & 88 \\
            \midrule
            $h$-$\phi$ Vanilla & 13.9k & 126.7 & 1.54\% & 0.936 & 4.9 & 5.9 \\
            $h$-$\phi$ FW & 13.9k & 126.7 & 1.55\% & 0.935 & 6.0 & 6.1\\
            $t$-$a$ SMSH & 24.2k & 124.4 & 0.35\% & 0.992 & 3.6 & 9.3 \\
            $t$-$a$ SMSH$^\dagger$ & 22.2k & 124.7 & 0.19\% & 0.992 & 2.6 & 8.6\\
            \midrule
            $h$-$\phi$ SMSH-TS & 28.2k & 125.4 & 0.51\% & 0.994 & 8.0 & 10.3 \\
            $h$-$\phi$ SMSH-TS$^\dagger$ & 24.7k & 125.5 & 0.55\% & 0.979 & 7.2 & 8.4 \\
            \bottomrule
      \end{tabular}
      \vspace*{0.5em}

      \footnotesize{$^*$single AMD EPYC Rome CPU-64cores at 2.9 GHz, using 8 cores.}
    \end{center}
    \vspace*{-1em}
\end{table}

Table~\ref{tab:numerical-performance} summarizes the quantitative comparison of all models, including accuracy and computational performance. The total time for assembling the linear systems and the total time required to solve them are denoted as $t_{\text{a}}$ and $t_{\text{s}}$, respectively. Note that different non-optimized assembly procedures are currently implemented for each model, so that assembly times cannot be compared directly. The SMSH models marked with~$^\dagger$ in Table~\ref{tab:numerical-performance} correspond to meshes with a single element through each homogenized bulk thickness.

The different detailed models give very similar results, both in terms of accuracy ($e_{P} < 1$\%, $R^2 > 0.988$) and solution time ($t_{\text{s}} \sim 90$~min). The $h$–$\phi$ formulation is considerably more efficient than its total $h$ counterpart, reducing the number of DoFs by almost half while producing identical physical results. The additional DoFs in the $h$-$\phi$ TS model compared to the $t$-$a$ model are due to the auxiliary fields within the thin-shells.

All homogenized models, including the $h$-$\phi$ SMSH-TS model, drastically reduce the problem size and the corresponding solution time while preserving accuracy. Although the $h$-$\phi$ Vanilla and FW models are the most efficient, they slightly overestimate the averaged AC losses. Moreover, SMSH models provide considerably more consistent current density distributions than the fully-homogenized models, with $R^2$ values comparable to that of detailed models other than the reference. While results with the $h$-$\phi$ Vanilla and FW models are almost identical, the FW model only requires the definition of a single cohomology function per stack~\cite{Denis2025}, which greatly facilitates its setup.

Notably, the $h$–$\phi$ SMSH-TS model remains accurate even when a single mesh element is used through each homogenized bulk, with a minimal degradation in $R^2$. Moreover, its computational performance is similar to that of the $t$–$a$ SMSH model, confirming the efficiency of the proposed approach. Overall, the results in Table~\ref{tab:numerical-performance} demonstrate that the $h$–$\phi$ SMSH-TS model recovers the accuracy of detailed formulations at roughly one tenth of their computational cost.

\begin{figure}[!t]
\centering
\includegraphics{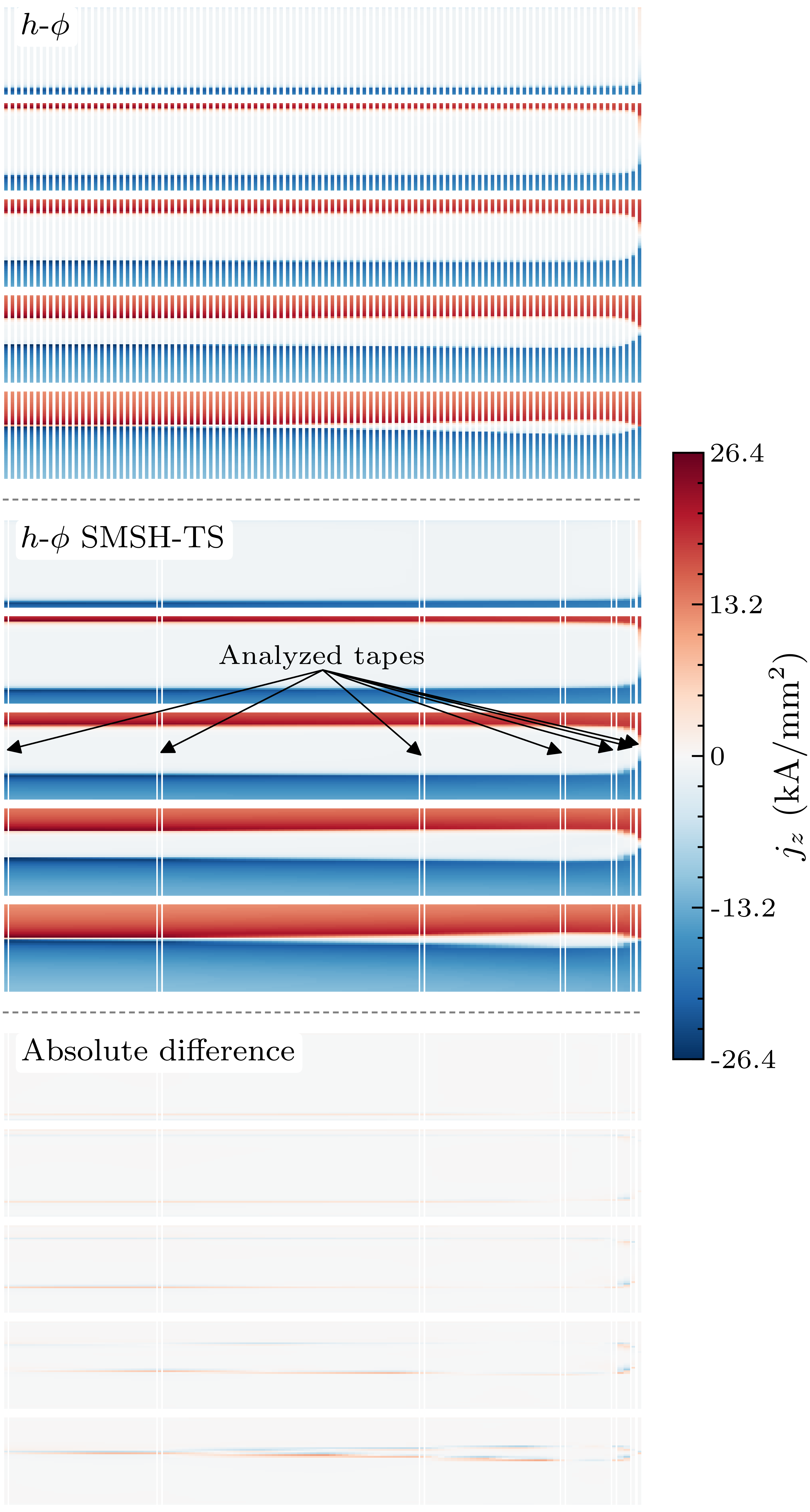}
\caption{Current density distribution at the second current peak ($t=3T/4$), computed with the reference $h$-$\phi$ model (top), the proposed $h$-$\phi$ SMSH-TS model (center) and their absolute difference (bottom). The current density in the homogenized bulks~\eqref{eq:js-def} is scaled by $(1/f_{\text{SC}})$ to ease comparison with the detailed model. The thickness of analyzed tapes is artificially enlarged to 160~$\upmu$m for better visibility.}
\label{fig:jlocal}
\end{figure}

Figure~\ref{fig:jlocal} compares the local current density distributions obtained at the second current peak. Again, the results show good agreement, with a high corresponding $R^2$ value of 0.994 (cf. Table~\ref{tab:numerical-performance}). The absolute difference between the two models is maximal (although still below $10$~kA/mm$^2$) at the front of current penetration, where the sharp transition from $\pm j_{\text{c}}$ to $0$ is not perfectly captured by the linear approximation~\eqref{eq:js-def}. Elsewhere, it remains consistent and effectively reproduces the background field generated by the non-analyzed tapes, leading to accurate loss estimations in the analyzed tapes.

As mentioned in Section~\ref{sec:smsh-ts}, AC losses in the non-analyzed tapes are obtained with the PCHIP method. Alternatively, instantaneous AC losses $q_{\text{s}}$ in homogenized bulks could be evaluated by integrating $\rho_{\text{s}} \lVert \j_{\text{s}} \rVert^2$ over $\Omega_{\text{s}}$, where $\rho_{\text{s}}$ follows the modified power-law:
\begin{equation}
    \rho_{\text{s}} = \frac{e_{\text{c}}}{f_{\text{SC}}~j_{\text{c}}} \left(\frac{\lVert \j_{\text{s}} \rVert}{f_{\text{SC}}~j_{\text{c}}} \right)^{n-1}. \label{eq:power-law-source}
\end{equation}
However, this alternative approach was found to overestimate the losses, with an averaged value of $P = 137.4$~W/m, against $P = 125.4$~W/m with the PCHIP method. This confirms the higher accuracy of the PCHIP method proposed in~\cite{Berrospe-Juarez2018} for predicting AC losses in the non-analyzed tapes.

\section{Conclusion \label{sec:5}}
The simultaneous multi-scale homogeneous (SMSH) method has been extended to the $h$-$\phi$ FE formulation. By furthermore collapsing the analyzed tapes into surfaces through the $h$-$\phi$ thin-shell (TS) model for easier mesh generation, this led to the proposed $h$-$\phi$ SMSH-TS model. Verification on a 2-D benchmark of 500 HTS tapes confirmed its accuracy for AC losses, turn voltages, and local current density. The method reproduces the accuracy of detailed $h$-$\phi$ and $t$-$a$ formulations at less than one tenth of the computational cost. This is similar to state-of-the-art homogenized models. Remarkably, it provides more reliable predictions of current density than fully-homogenized models, making it particularly well-suited to simulating more localized phenomena such as local defects in HTS tapes, among others.

Future work targets the application of the $h$-$\phi$ SMSH-TS model to 3-D simulations. Because it relies on the scalar magnetic potential $\phi$ outside analyzed tapes, the proposed approach is expected to provide a significant computational advantage over total $h$ and $t$-$a$ based approaches. Moreover, extending the model to no-insulation coils should be investigated.

\section*{Acknowledgments}
The authors gratefully acknowledge Sebastian Schöps (TU-Darmstadt), Elias Paakkunainen (TU-Darmstadt, Tampere University) and Paavo Rasilo (Tampere University) for the fruitful collaboration on the $h$-$\phi$ foil-winding method.

\balance

\bibliography{eucas_2025_hts}

\end{document}